\documentclass[10pt,letterpaper]{article}
\usepackage{opex3}
\usepackage{color}

\usepackage{cite}
\usepackage{amsmath}

\usepackage{bm}

\begin{document}

\title{Photo-induced voltage in nano-porous gold thin film}
\author{Marjan Akbari,$^1$ Masaru Onoda$^2$ and Teruya Ishihara $^{1,*}$}
\address{$^1$Department of Physics, Tohoku University, Sendai, 980-8578, Japan\\
$^2$Department of Electrical and Electronic Engineering, 
Akita University, Akita, Japan}

\email{$^*$t-ishihara@m.tohoku.ac.jp}

\begin{abstract}
\noindent We report an experimental study of generation of photo-induced voltage
in nano-porous gold ({\small NPG}) thin film under
the radiation of obliquely incident nanosecond laser light in visible
regions. For s- polarized light, negative voltage is
observed along the incident plane for positive incident angles, while  
for p- polarized light, positive  voltage is observed for wavelength
longer than 510 nm, while it turns to negative for shorter wavelengths.
The transverse voltage for various polarized light is explained in terms of symmetry of configuration and that of microscopically random but  macroscopically isotropic {\small NPG}.
\end{abstract}



\section{Introduction}

It is known that photo-induced voltage is generated in conducting thin film during the interaction of laser pulses and the simplest description for this observed voltage is that when structure is radiated by light, light can push electrons through Lorentz force due to its electric and magnetic fields. The force on electrons can be detected as a voltage in the structure. In simple cases, this force can be estimated using momentum conservation, that is, momentum from photons transfers to the free carriers of the material and this effect manifests itself as a pulsed voltage. This phenomenon has been investigated by other groups in various materials and explained in terms of optical rectification and photon drag effect.
 
In systems without inversion symmetry, light can induce DC polarization in the structure that is second order in terms of electric field. This term is referred to as optical rectification and is responsible for the voltage. Photo-induced voltages were reported for metallic artificial structures with asymmetric unit cells when laser beam is normally incident to the sample \cite{Hatano1, Kurosawa2}. 

Although photo-induced voltage for oblique incidence of light can be treated as a case without inversion symmetry, a term of photon drag effect is often used. This effect is significant in semiconductors \cite{Luryi1987} and very small in bulk metals\cite{Gurevich2000}.  The induced current due to photon drag effect in metals has been described using equation of motion for the electrons within a hydrodynamic model\cite{Goff}. Enhancement due to surface plasmon excitation has been observed with Kretschmann configuration \cite{Vengurlekar2005} as well as grating coupling \cite{Kurosawa1}.  It has been reported that the transverse photo-induced voltage is generated perpendicular to the incident plane for circularly and $45^\circ$  tilted linear polarization in plasmonic crystal slabs \cite{Hatano2009}. 
Similar phenomena were observed in some conductive thin films \cite{Mikheev2, Mikheev3}, where the definition for transverse and longitudinal configurations is opposite to ours.  Plasmon drag effect in gold quasi-periodic nanomesh structure was compared to flat film of gold \cite{Noginova2} and it showed strong voltage in the former where its peak of voltage was near localized plasmon resonance. The results show that geometry of nanostructure can control polarity and magnitude of the voltage.

In this paper we report for the first time photo-induced voltage in Nano-Porous Gold ({\small NPG}) thin film, which is a random nanostructure. {\small NPG} is generated by corrosion of an alloy of gold and less noble material such as silver. It has attracted attention because of simple preparation techniques to fabricate nanostructures compare to other methods such as Electron Beam (EB) or Fast Ion Beam (FIB) lithography. {\small NPG} is used in the field of sensors, actuators, optics, and many more areas \cite{wittstock2010}. 

Because of these nano-size pores in the structure, surface plasmon-like excitation is expected upon light irradiation. Conductive character allows detection of photo-induced voltage in {\small NPG}. We have studied polarity and amplitude of the pulsed response voltage of {\small NPG} and show that they depend on the incidence angle, type of polarization and wavelength of incident light.

\section{Experimental}

\begin{figure}[t]
 \begin{center}
      \includegraphics[width=\linewidth]{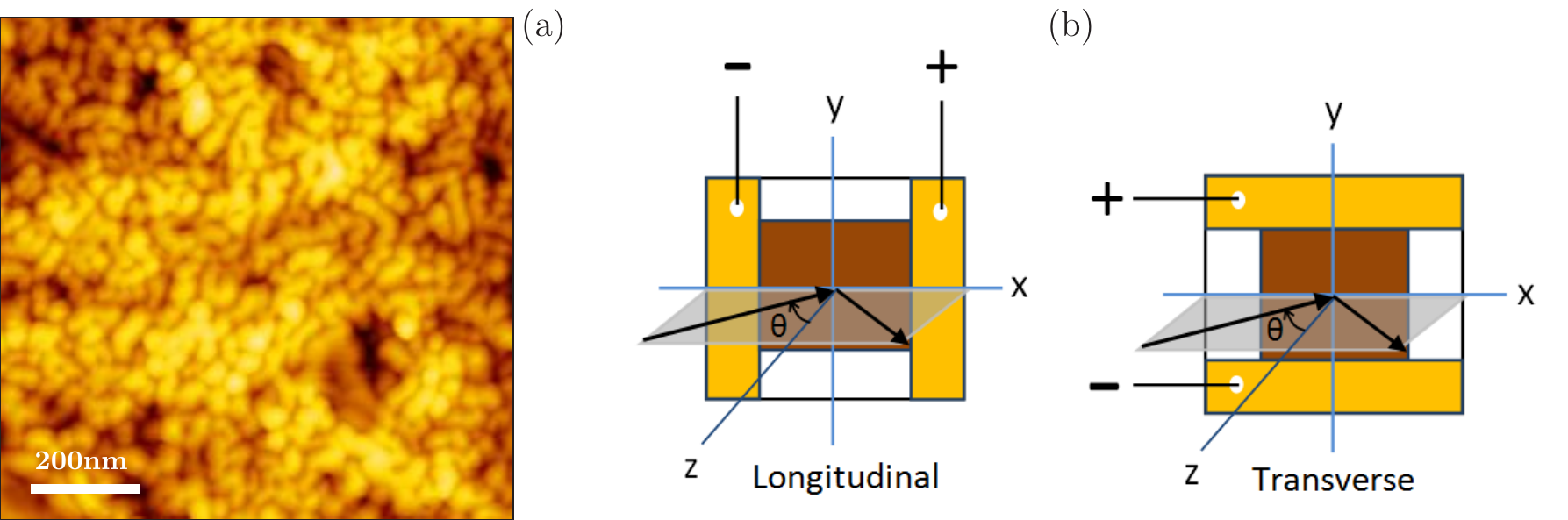}

        \caption{\label{Configs}(a) AFM image of \small{NPG} film shows that it is a network of pores and gold. (b) Two configurations for measuring PIV in {\small NPG}. The arrow shows definition of the positive incident angle ($\theta$) in the configurations.}

 \end{center}
\end{figure}

{\small NPG} thin film can be generated by dealloying of some suitable alloys of gold like AuAg, AuCu and AuNi. In this process, the less noble component is removed and the more noble material in the alloy, here gold, remains and forms an interconnected nanostructure of pores and gold (Fig. \ref{Configs}(a)). We dissolved silver from alloy of 50$\%$ Au 50$\%$ Ag, which is commercially available with 100 nm thickness, by exposure to 65$\%$ concentrate $\mathrm{HNO}_3$ for 30 minutes. Then it was taken out from acid and rinsed in distilled water to remove the residual acid. After rinsing, the film floating on water surface was transferred onto a drop of water at the center of a 15 $\times$ 15 mm glass substrate with evaporated Au electrodes. The drop of water was removed afterwards. It is convenient to use a piece of filter paper to transfer the film. Then the film was trimmed with a focused laser beam into of dimensions 3 $\times$ 3 mm. Typical thickness of the {\small NPG} films was about 100 nm.

Each sample was provided with two parallel electrodes. For connecting two electrodes to the  {\small NPG} film for measuring photo-induced voltage, we used vacuum evaporator to make a 50 nm layer of gold on the glass substrate with a gap for attaching  {\small NPG} film.  Before that, a 5 nm thick intermediate layer of Cr  was evaporated on the substrate for better adhesion. Then {\small NPG} film was directly mounted in the gap, and pieces of copper tapes were connected on the evaporated gold to make connection between film and electrodes. Finally, two leads as electrodes were soldered to the copper tapes. After the attachment of the electrodes, the substrate was fixed on a 20 $\times$ 20 mm printed circuit board (PCB). The typical ohmic resistance of the films was 12 $\Omega$.

In our experiment, we employed two configurations; longitudinal (the electrodes were oriented perpendicular to the incidence plane and the voltage along x axis was measured) and transverse (the electrodes were oriented parallel to this plane and the voltage along y axis was measured) (see Fig. \ref{Configs}(b)). The observed photo-induced voltages; longitudinal photo-induced voltage ({\small LPIV}) and transverse photo-induced voltage ({\small TPIV}), were measured for different types of polarization (circular polarized light and $\pm45^\circ$ linear polarized light for transverse configuration. Circular polarized light, s- polarized light and p- polarized light for longitudinal configuration). Positive incident angle is shown by the arrow in Fig. \ref{Configs}(b). For angle resolved measurement the laser beam is fixed in the xz plane and by rotating the film on a rotational stage around y axis of the film, incidence angle was varied between $-85^\circ$ to $+85^\circ$ by  at an interval of $1^\circ$. 

When free electrons are dragged toward positive electrode with the projection of k vector in that direction (direction of the photo-induced voltage is toward the positive electrode of the sample), we have a negative signal on the monitor of the oscilloscope. Peak of this negative signal is measured as a negative voltage in angle resolved and wavelength resolved voltages and in the graph of data it is shown with a negative voltage.

\begin{figure}[tp]
    \begin{center}
      \includegraphics[width=\linewidth]{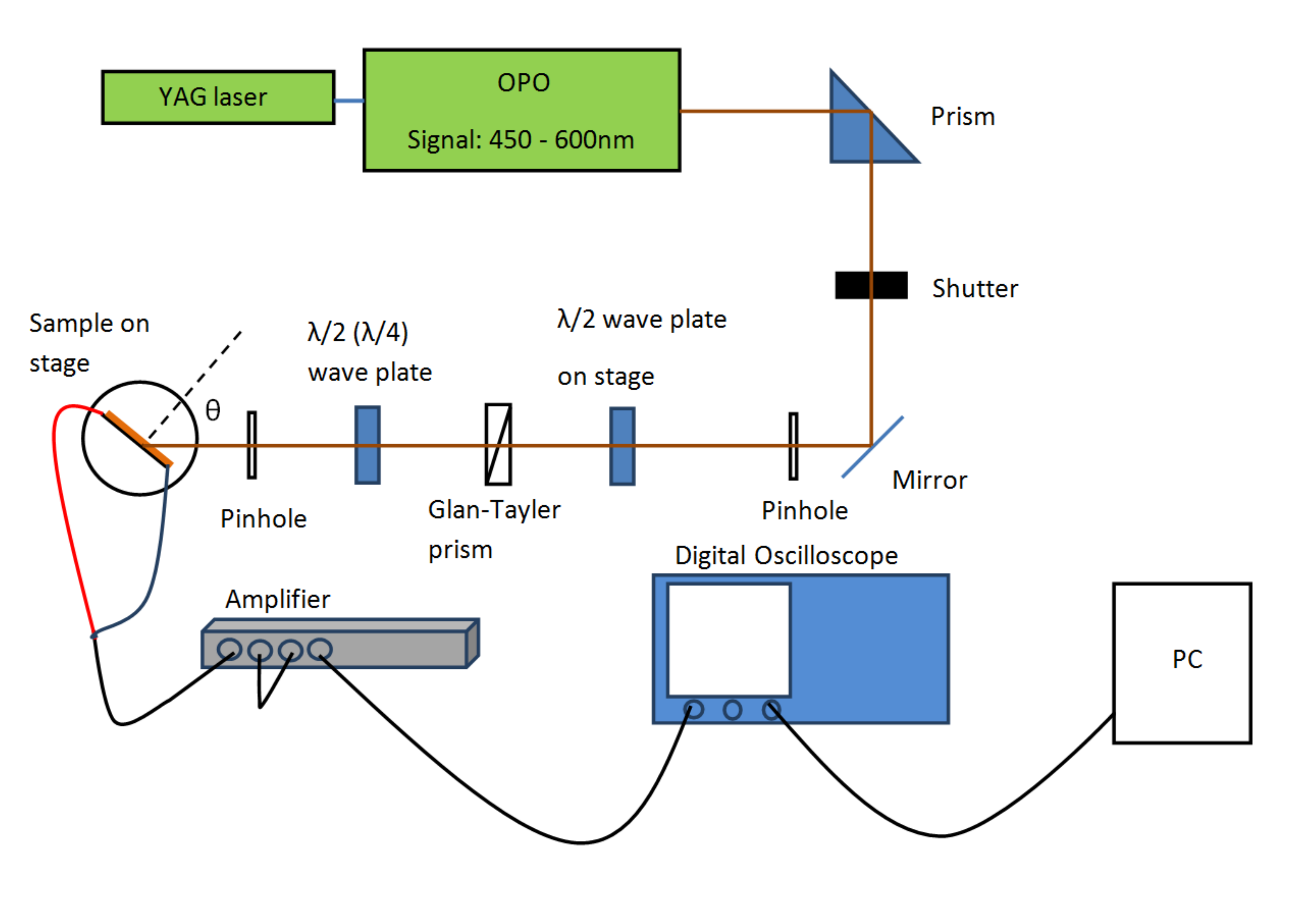}  
    \end{center}
    \caption{\label{PIVsetup}Setup of the measuring photo-induced voltage in {\small NPG} film.}
\end{figure}

\begin{figure}[tp]
\footnotesize
   \begin{center}
 \includegraphics[width=0.7\linewidth]{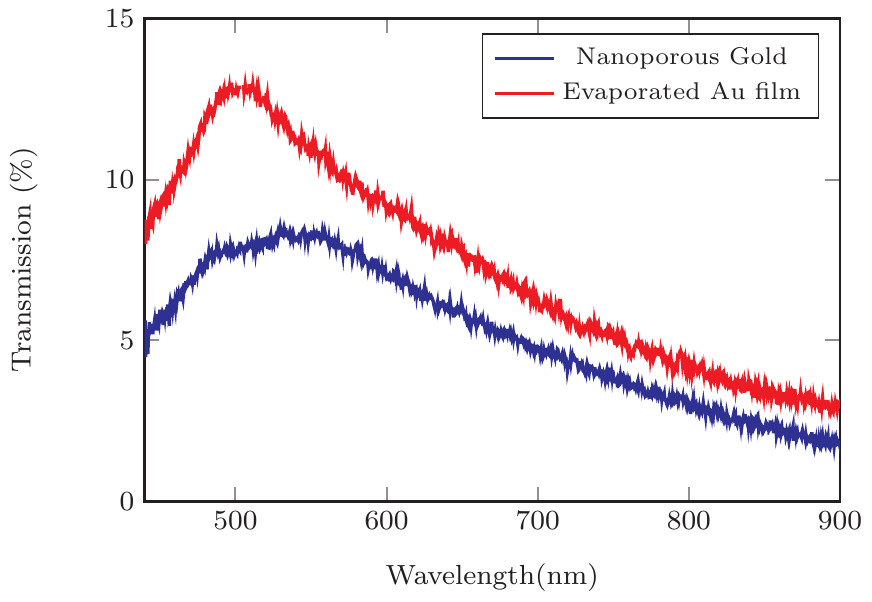}
\end{center}
     \caption{\label{RTdata}Transmission spectra for {\small NPG} (blue line) and evaporated gold (red line).}
\end{figure}

A schematic diagram of the experimental setup is shown in Fig. \ref{PIVsetup}. Light from an optical parametric oscillator ({\small OPO}) was pumped by tripled Nd:YAG laser and produced both idler wavelengths (light with 710-2300 nm) and signal wavelengths (light with 450-700 nm).
The laser pulse width and repetition rate were 7 ns and 10 Hz, respectively. The polarization of light was controlled with an achromatic half-wave plate or quarter-wave plate from Thorlabs. The beam size was chosen so that it covers the sample at the normal incidence. For keeping intensity of incident light constant (4 MW/cm$^2$ at normal incidence for all wavelengths) we used a half-wave plate on a rotational stage between {\small OPO} and a Glan-Tayler prism in the setup. Laser pulse energy for this intensity was 0.3 mJ and threshold of the laser damage intensity was 24 MW/cm$^2$.

A sample was mounted on a rotational stage, so that we can change incidence angle of the light for measuring angle resolved photo-induced voltage. We measured peak of photo-induced voltage for different configurations with a digital oscilloscope (Tektronix TDS3012B with 50 $\Omega$ input impedance and pass band of 100 MHz) which is triggered by Q-switch of the laser. An amplifier with a gain of 125 was used before feeding induced voltage to the oscilloscope. For reducing pulse to pulse fluctuations in the photo-induced signal, they were averaged for 32 pulses.

\section{Experimental results}

\subsection{Transmission Spectrta}

In order to characterize our sample, we first compared transmission spectrum of {\small NPG} film with an evaporated gold film. As is shown in Fig. \ref{RTdata}, transmission peak of the {\small NPG} film locates significantly lower than that of the evaporated gold film,  which is due to the lower effective electron density and responsible for the copper-like color of {\small NPG}.

\subsection{Transverse voltage}

We measured photo-induced voltage along y axis of the {\small NPG} film in transverse configuration with circularly polarized and $\pm45^\circ$ linear polarized light. 

As we see in Fig. \ref{Configs}(b) {\small NPG} film is in the xy plane and incident laser beam is in the zx plane. In this condition with circular polarized light, when electric field of the light rotates clockwise in the xy plane as is looked toward the propagation direction, we refer this polarization as Right Circular Polarization  ({\small RCP}) and when it rotates anticlockwise, we define it as Left Circular Polarization ({\small LCP}).

\begin{figure}[tbp]
\footnotesize
\includegraphics[width=\linewidth]{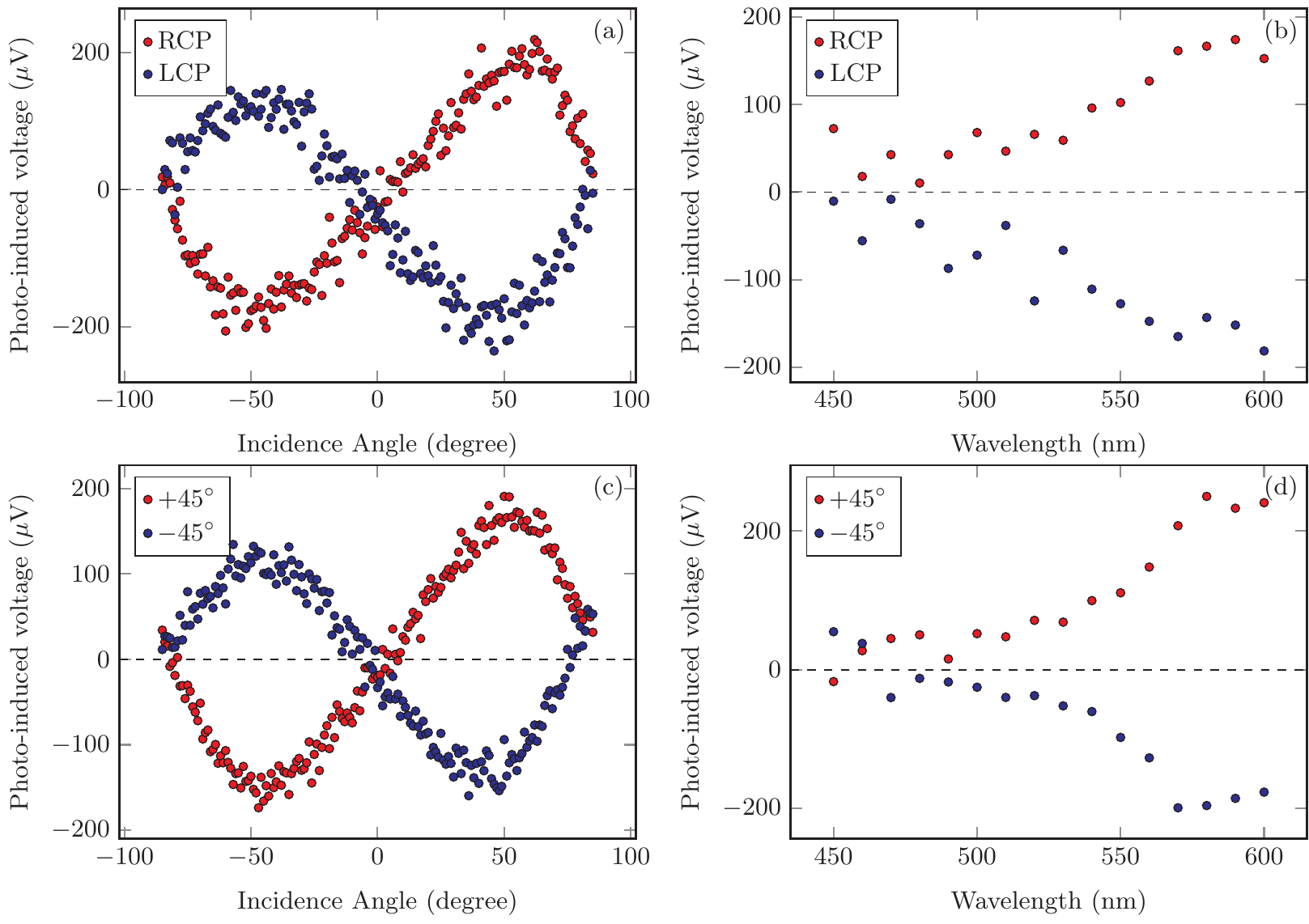}
\caption{\label{TPWCIR}(a) Angle resolved TPIV for 550 nm, circular polarized light. (b) Wavelength resolved TPIV for  $+50^\circ$ incidence angle, circular polarized light. (c) Angle resolved TPIV for 550 nm, $\pm45^\circ$  linear polarized light. (d) Wavelength resolved TPIV for $+50^\circ$ incidence angle, $\pm45^\circ$  linear polarized light.}
\end{figure}

Shape of the signal resembles shape of laser pulses and is proportional to power of the applied light. Figure \ref{TPWCIR}(a) presents the experimental dependence  of the photo-induced voltage on the incidence angle, $\theta$, for transverse configuration with {\small RCP} and {\small LCP}.

For the same incidence angle, sign of the voltage for {\small RCP} is opposite of voltage for {\small LCP}. So, polarity of the voltage changes as the sense of polarization rotation changes from {\small RCP} to {\small LCP}. By increasing incidence angle from zero to larger angles, voltage increases for both negative and positive directions. At about $\theta$= $+50^\circ$, for {\small RCP} there is a maximum positive voltage. For {\small LCP}, we have a maximum negative voltage at this angle. For angles larger than $\theta$=$+50^\circ$ voltage decreases and at $\theta$=$\pm85^\circ$  voltage is almost zero. So, angle dependency of amplitude of the voltage can be proportional to $\sin(2\theta)$ for both {\small RCP} and {\small LCP}, which is proportional to the amount of photon momentum along the sample multiplied with photon flux through the sample.

Figure \ref{TPWCIR}(b) shows dependence  of the photo-induced voltage on the wavelength, $\lambda$, for transverse configuration with {\small RCP} and {\small LCP} light at fixed incidence angle, $\theta$= $+50^\circ$. As it is clear from this figure, intensity of voltage between 450 to 520 nm is almost the same but for longer wavelengths it gradually increases. 

Sign of the TPIV does not change by changing wavelength in this wavelength range, for {\small RCP} it is always positive and for {\small LCP} always negative. For producing $\pm45^\circ$ linear polarized light, an achromatic quarter-wave plate was replaced with the half wave plate in the setup. For both cases the Glan-Tayler prism was adjusted for p- polarization. Figure \ref{TPWCIR}(c) shows experimental dependence  of the photo-induced voltage on the angle of incidence for transverse configuration with $\pm45^\circ$ linear polarized light. For both transverse photo-induced voltage with circular and $\pm45^\circ$ linear polarized light at incidence angle of zero there is a small nonzero voltage that could be due to inhomogeneities in the film.

By increasing incidence angle from zero to larger angles, voltage increases for both negative and positive directions. Sign of the voltage for $+45^\circ$ linear polarization is opposite of the $-45^\circ$ linear polarization. So, polarity of the voltage changes as the sense of polarization changes from positive to negative.

At about $\theta$=$+50^\circ$, for $+45^\circ$ linear polarization, there is a maximum positive voltage. For $-45^\circ$, we have a maximum negative voltage at this angle. For angles larger than $\theta$=$+50^\circ$ voltage decreases and at $\theta$=$\pm 85^\circ$  voltage vanishes.

Figure \ref{TPWCIR}(d) shows dependency  of the photo-induced voltage on the wavelength for transverse configuration with $\pm45^\circ$ linear polarized light. It shows  wavelength dependence of voltage is similar to that for circular polarization as we saw above (intensity of induced voltage between 450 to 520 nm is almost the same but for larger wavelengths it gradually increases). Sign of the TPIV does not change by changing wavelength in this wavelength range (for $+45^\circ$ it is always positive and for $-45^\circ$ always negative).

\subsection{Longitudinal voltage}

\begin{figure}[tbp]
\footnotesize
  \includegraphics[width=\linewidth]{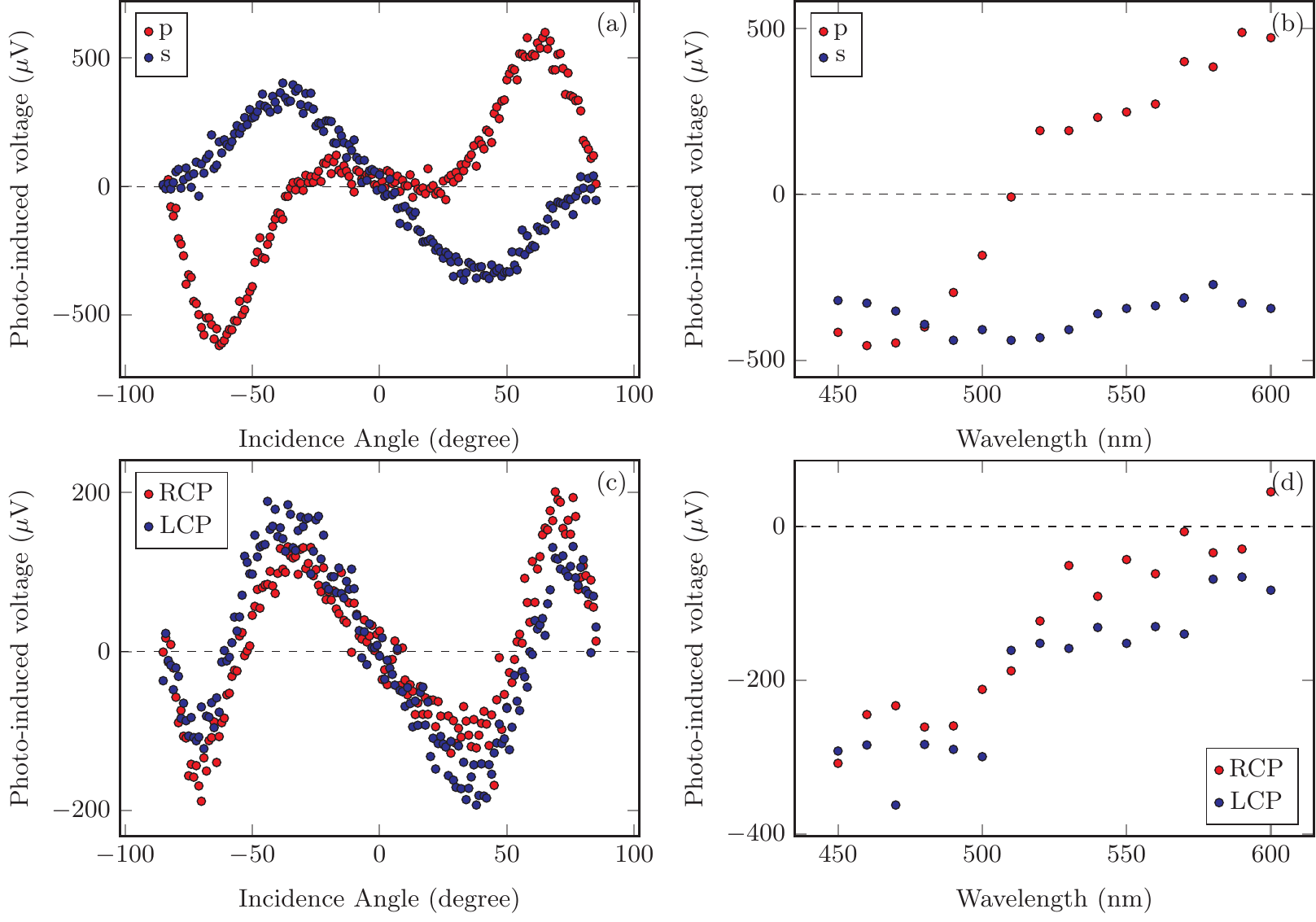} 
\caption{\label{LPSP}(a) Angle resolved LPIV for 580 nm, p- and s- polarized light. (b) Wavelength resolved LPIV for $+50^\circ$ incidence angle, p- and s- polarized light. (c) Angle resolved LPIV for 580 nm, circular polarized light. (d) Wavelength resolved LPIV for $+50^\circ$ incidence angle, circular polarized light.}
\end{figure}
  
We measured photo-induced voltage along x axis of the film in longitudinal configuration with circularly polarized and s- and p- polarized light. By rotating sample $+90^\circ$ around z axis, configuration of the film changes from transverse to longitudinal. For producing p- and s- polarized light with 550 nm, a Glan-Tayler prism and half wave plate were used in the setup. 

Angle resolved photo-induced voltage at the wavelength of 550 nm in longitudinal configuration with s- and p- polarized light are shown in the Fig. \ref{LPSP}(a). For s-polarized light it follows $\sin(2\theta)$ dependence and the sign is negative for positive angles. It is reasonable as electron has negative charge, which is pushed by light momentum. At about $\theta$=$+50^\circ$, for s- polarization there is a maximum negative voltage. Surprisingly for p-polarization, photo-induced voltage is almost zero upto $\theta$=$+30^\circ$ and becomes positive for larger positive incident angle. At about $\theta$=$+65^\circ$ there is a maximum positive voltage. 

Wavelength resolved photo-induced voltage in longitudinal configuration with p- and s- polarized light are shown in Fig. \ref{LPSP}(b). For s- polarization, it is always negative for positive incident angles. On the other hand, we found that for p- polarization the sign of the voltage changes depending on the wavelength, from 450 to 510 nm this voltage is negative, while from 510 to 600 nm it is positive.

Angle dependence of photo-induced voltage in longitudinal configuration with {\small RCP} and {\small LCP} are shown in Fig. \ref{LPSP}(c). Sign and intensity of the voltage for {\small RCP} is the same with {\small LCP}. So, voltage does not change as the sense of circular polarization changes from {\small RCP} to {\small LCP}. At about $\theta$=$+70^\circ$, for both {\small RCP} and {\small LCP} there is maximum positive voltage. At $\theta$=$-70^\circ$ there is  maximum negative voltage. For angles larger than $\theta$=$\pm70^\circ$ voltage decreases and at $\theta$=$\pm85^\circ$  there is no voltage. It seems this voltage is made by adding s- and  p- polarization together. Wavelength dependence photo-induced voltage in longitudinal configuration with {\small RCP} and {\small LCP} are shown in Fig. \ref{LPSP}(d) for $\theta$=$+50^\circ$. This data suggests that  wavelength dependence for circular-polarized  light excitation is the average of curves for s- and p-polarized light in Fig. \ref{LPSP}(b) as is consistent with the angle dependence in Fig. \ref{LPSP}(a) and 5(c).

Both longitudinal photo-induced voltage with linear polarized light (s- and p-) and circular polarized light were measured with the same sample and configuration and position of laser light on the sample. The size of light was small enough to have the least possible of light on the electrodes. This complicated tendency in Fig. \ref{LPSP}(c) cannot be due to the irradiation of the electrodes by light. If it would be so, we should have seen this possible effect on the measured data at around larger incident angles in Fig. \ref{LPSP}(a).

\section{Discussion}

As we saw in all measurements, photo-induced signal intensities are following nanosecond laser pulses proportional to the power of the laser beam. 
Amplitudes of the signals are proportional to a characteristic law, $\sin(2\theta)$, 
for all TPIV and s-polarization case for LPIV, which is proportional to photon momentum and photon flux on a finite area of the sample.
There are no TPIV for s- or p- polarized light. 

The observed tendencies of angle resolved photo-induced voltage in the {\small NPG} thin film are generally
similar to the experimental measurements of angle resolved voltages for thin films of nanocarbon film \cite{Mikheev4},  nanographite film \cite{Mikheev1} and single-walled carbon nanotubes \cite{Mikheev3}. In these films positive voltage is generated along the sample for positive incident angle, which is consistent with the positive carriers in these materials.   In our experiment, we found that the sign of the photo-induced voltage for positive incident angles is negative for s-polarized light, which is expected for Au. Surprisingly as we saw in Fig. \ref{LPSP}(b), voltage sign for p-polarized light is wavelength dependent: it is positive for wavelengths longer than 510 nm, while it is negative for those shorter than that. Note that we observed the second peak in transmission spectrum in Fig. \ref{RTdata}. This spectral region was assigned as localized surface plasmon resonance in {\small NPG}\cite{Lang} as it shifts with the average pore size. This surface plasmon polariton-like excitation might be responsible for the observed opposite voltage sign in wavelength resolved voltage for p- polarization (red data points in
Fig. \ref{LPSP}(b)).

Noginova et al. have investigated photon drag effect in gold and silver nanostructures  \cite{Noginova2}. They observed that electrons are always  dragged in the direction of its wavevector for the wavelength they investigated (420-750 nm). It is interesting to examine similar systems with nanostructure smaller than that in \cite{Noginova2} but larger than that of the {\small NPG} in the present study, which might clarify the origin of the sign change in {\small NPG}.

\begin{figure}[tbp]
        \includegraphics[width=\linewidth]{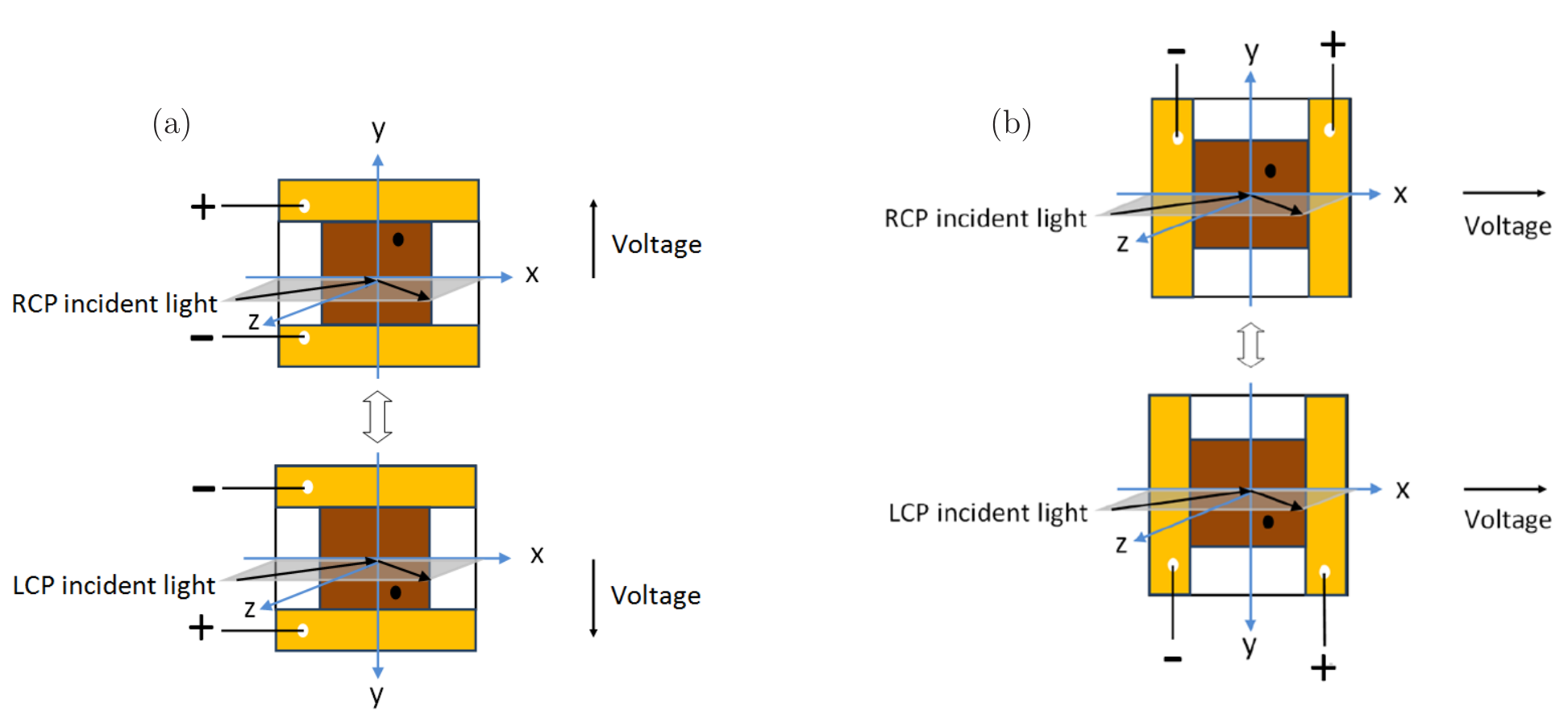}       
    \caption{\label{TLmirr}(a) Mirrored transverse configuration. (b) Mirrored longitudinal configuration.}
\end{figure}

As for TPIV, it is a characteristic feature that the voltage change its sign by changing the sense of circular polarization or $\pm45^\circ$ linear polarization. 
In case of plasmonic crystal slabs, it is possible to explain these responses by means of the symmetry of field in a unit cell \cite{Hatano2009}. 
In order to explain the polarization dependence
of photo-induced voltage in such a random structure as {\small NPG},  we need more general reasoning.
For this purpose, let us consider the mirror transformation ($y\leftrightarrow -y$)
of the whole system including incident beam
with respect to the incident $xz$-plane
as Fig. \ref{TLmirr}(a) for TPIV and Fig. \ref{TLmirr}(b) for LPIV.
First, we introduce the response function of photo-induced voltage as

\begin{eqnarray}
\bm{V}\left(\bm{E}\right) &=&
\begin{pmatrix}
V_{x}\left(\bm{E}\right) \\
V_{y}\left(\bm{E}\right) 
\end{pmatrix}
\end{eqnarray}

for the case of incident electromagnetic wave with electric field $\bm{E}$.
Through the transformation depicted in Fig. \ref{TLmirr}(a),
we can easily see that the following relation exactly holds
between the response of the original system and that of its mirror image, $\bm{V}^{\mathrm{m}}$,

\begin{eqnarray}
\bm{V}^{\mathrm{m}}\left(\bm{E}^{\mathrm{m}}\right) &=&
\begin{pmatrix}
1&0 \\
0&-1
\end{pmatrix}
\bm{V}\left(\bm{E}\right) 
=
\begin{pmatrix}
V_{x}\left(\bm{E}\right)\\
-V_{y}\left(\bm{E}\right) 
\end{pmatrix}
\end{eqnarray}

where $\bm{E}^{\mathrm{m}}$ is the mirror image of $\bm{E}$.
On the other hand, when a {\small NPG} thin film is too random to be
macroscopically distinguished from its mirror image,
the relation $\bm{V}\left(\bm{E}\right)\cong \bm{V}^{\mathrm{m}}\left(\bm{E}\right)$
should be approximately satisfied,
and leads to the phenomenological consequence.

 \begin{eqnarray}
\begin{pmatrix}
V_{x}\left(\bm{E}^{\mathrm{m}}\right) \\
V_{y}\left(\bm{E}^{\mathrm{m}}\right) 
\end{pmatrix}
&\cong&
\begin{pmatrix}
V_{x}\left(\bm{E}\right) \\
-V_{y}\left(\bm{E}\right) 
\end{pmatrix}
\end{eqnarray}\\

This argument explains the polarity with respect to the polarization state of incident beam
for every angle resolved measurement in Fig. \ref{TPWCIR}(a) and 4(c) and Fig. \ref{LPSP}(a).

Next, bringing in a hypothesis, we would like to push forward the above discussion
to predict photo-induced signals for generic polarization through the data in hand.
Our proposal for the hypothesis is summarized as follows,

\begin{eqnarray}
\bm{V}\left(\bm{E}\right) 
&=&
\begin{pmatrix}
\left|z_{\mathrm{p}}\right|^{2}V_{\mathrm{p}}\left(\left|\bm{E}\right|,\theta\right)
+\left|z_{\mathrm{s}}\right|^{2}V_{\mathrm{s}}\left(\left|\bm{E}\right|,\theta\right)
\\
2\left[
\mathrm{Re}\left(z^{*}_{\mathrm{p}}z_{\mathrm{s}}\right)V_{\mathrm{d}}\left(\left|\bm{E}\right|,\theta\right)
+\mathrm{Im}\left(z^{*}_{\mathrm{p}}z_{\mathrm{s}}\right)V_{\mathrm{c}}\left(\left|\bm{E}\right|,\theta\right)
\right]
\end{pmatrix}
\end{eqnarray}

where 
$\bm{E} = \left|\bm{E}\right|
\left(z_{\mathrm{p}}\bm{e}_{\mathrm{p}}+z_{\mathrm{s}} \bm{e}_{\mathrm{s}}\right)$
, and
$z_{\mathrm{p}}$ and $z_{\mathrm{s}}$ 
are complex coefficients satisfying 
$\left|\bm{z}\right|^{2}=\left|z_{\mathrm{p}}\right|^{2}+\left|z_{\mathrm{s}}\right|^{2} = 1$
and representing the polarization state of incident beam,
$\bm{e}_{\mathrm{p(s)}}$ is the polarization vector of p(s)- polarization,
and $\theta$ is the angle of incidence.
By appropriately choosing the functions $V_{\mathrm{p,s,c,d}}$,
we can make this form of response consistent with all the data in hand.
Reversely, the consistency of this hypothesis with the data in Fig. \ref{LPSP}(a)
fixes $V_{\mathrm{p}}$ and $V_{\mathrm{s}}$, 
and the consistency with the data in Fig. \ref{TPWCIR}(a) and 4(c) fix $V_{\mathrm{c}}$ and $V_{\mathrm{d}}$ respectively.
The results in Fig. \ref{TPWCIR}(a) and 4(c) also suggest $V_{\mathrm{c}}\cong V_{\mathrm{d}}$
at least for the case of $\lambda = 550$ nm. 
Further analysis on the verification and validation of the hypothesis
will be reported elsewhere.

\section{Conclusions}

We have experimentally investigated angle and wavelength dependence of TPIV and LPIV in {\small NPG} thin film.
The $\sin (2\theta)$ dependence observed in TPIV for circular polarized light and LPIV for s- polarized light suggests photon momentum transfer to the sample. On the other hand, LPIV shows extraordinary response when the excitation involves p- polarization for wavelength longer than 510 nm:  The photo-induced voltage is positive for positive incident angle, which cannot be explained by simple momentum transfer argument. In order to connect this behavior to surface plasmon like excitation, measurement of permittivity in {\small NPG} is in progress.\\

\textbf{Acknowledgments}

This work was partly supported by Research Foundation for Opto-Science and Technology and Grant-in-Aid for Scientific Research on Innovative Areas ”No.22109005” from The Ministry of Education, Culture, Sports, Science and Technology (MEXT), Japan.

\end{document}